\documentclass[12pt]{article}

\usepackage{sbc-template}
\usepackage{graphicx,hyperref}
\usepackage[utf8]{inputenc}
\usepackage[brazil]{babel}
\usepackage{xcolor}
\usepackage{float}
\usepackage{multirow}

\title{Blended Integrated Open Data: dados abertos públicos integrados}

\author{Fabiola Santore,
Lucas F. Oliveira, Rafael de Paulo Dias,
\\ Henrique V. Ehrenfried, Alessandro Elias, Diego Pasqualin,
\\Luis C. E. de Bona, Marcos Didonet Del Fabro, Marcos Sunyé}

\address{Centro de Computação Científica e Software Livre (C3SL) \\
Departamento de Informática \\
Universidade Federal do Paraná (UFPR)\\
R. Cel. Francisco H. dos Santos, 100 -- Curitiba -- PR -- Brasil
\email{\{fsantore,lfoliveira,rpd17,hvehrenfried,aelias\}@inf.ufpr.br}
\email{\{dpasqualin,bona,marcos.ddf,sunye\}@inf.ufpr.br}
}

\newcommand{\blendb}{\emph{BlenDB}}

\newcommand{\simmc}{SIMMCTIC}
\newcommand{\api}{http://biod.c3sl.ufpr.br/api/v1}

\begin{document} 

\maketitle

\begin{abstract}
  While several public institutions provide its data openly, the effort required to access, integrate and query this data is too high, reducing the amount of possible dataset users. The Blended Integrated Open Data (BIOD) project has as objective to ease the access to public Open Data. It integrates and makes available more than 300Gb of data, containing billions of records from different Open Data Sets, allowing to query over them, and thus to retrieve related information from originally disconnected data sets. This paper presents the set of open data available, how to access it and how produce new compatible data to improve the existing data set.
\end{abstract}
     
\begin{resumo} 
  Embora diversas instituições públicas disponibilizem dados abertamente, o esforço necessário para acessar, integrar e consultar esses dados é muito grande, reduzindo a quantidade de possíveis usuários da base de dados. O projeto \emph{Blended Integrated Open Data} (BIOD) tem objetivo de facilitar o acesso a dados abertos públicos. O projeto integra e disponibiliza mais de 300 Gb de dados, constituído por bilhões de registros extraídos de diferentes fontes abertas, permitindo e facilitando a consulta, possibilitando a recuperação de informações relacionadas em bases originalmente desconectadas. Este artigo apresenta o conjunto de dados abertos disponível, como acessá-lo e como produzir novos dados abertos compatíveis para serem incluídos no repositório existente.
\end{resumo}

\section{Introdução}\label{sec:intro}

A quantidade de dados abertos disponíveis, principalmente fornecidas por instituições públicas, é abundante, no entanto, não existe um padrão de divulgação de dados abertos nem uma ferramenta que permita encontrar essas bases facilmente. Existem diversos portais governamentais com grande quantidade de dados e microdados abertos; citamos como exemplo o \textit{Portal Brasileiro de Dados Abertos}, ou o \textit{Portal do INEP com microdados educacionais}, que disponibilizam dados abertos que permitem a criação de indicadores importantes. Entretanto, o cruzamento de diferentes bases de dados é uma tarefa complexa, difícil de ser implementada, pois é necessário realizar um processo de integração e manutenção ao longo do tempo. Desta forma, esses dados abertos acabam sendo subutilizados.

O projeto \emph{Blended Integrated Open Data} foi criado pelo grupo C3SL \cite{c3sl2016} da Universidade Federal do Paraná, para tornar os dados abertos mais acessíveis e aumentar a utilização de grande massa de dados abertos. Para atingir esse objetivo, o projeto criou um repositório composto de diversas bases de dados abertas distintas e utilizou uma ferramenta que permite o acesso integrado a essas bases de dados. Além da disponibilização das bases de dados, essas bases foram padronizadas e integradas, com a finalidade de aumentar ainda mais a usabilidade desses dados.


O foco do projeto não é apenas a disponibilização de uma base de dados mas também a sua utilização de forma integrada, com a possibilidade de realização consultas com cruzamentos complexos de dados através de API (Application Programming Interface) de consulta. Pensando na utilização desses dados para consultas do tipo OLAP, além da disponibilização dos ``microdados'', os dados do repositório também podem ser obtidos de forma pré-agregada. A pré-agregação pode reduzir consideravelmente massa de dados \cite{kimball2011data}, o que reduz a quantidade de recurso para processar os dados, seja tempo, espaço de armazenamento ou tráfego de rede, o que pode gerar uma análise mais rápida, a custo de precisão na análise.

Atualmente o repositório está disponível abertamente \footnote{Repositório de dados abertos: \url{https://biod.c3sl.ufpr.br/}} e 
possui dados abertos sobre: escolaridade, economia, população e utilização de internet por programas de inclusão digital brasileiros. São 24 tabelas diferentes, mais de 900 atributos e quase 3 bilhões de registros armazenados e integrados de forma inédita. Além da manutenção do repositório, isto é, atualização do conteúdo do repositório conforme mais dados são liberados pelas fontes originais, espera-se que esse repositório seja um dos primeiros passos na criação de um repositório colaborativo de dados abertos.


\section{Descrição da base de dados}\label{sec:conteudo}
Os dados disponíveis no repositório de dados integrados foram extraídos de outras bases de dados abertos. Os dados foram extraídos das bases de dados dos projetos \simmc{} e LDE.


O \simmc{}\footnote{SIMMCTIC: \url{https://simmctic.c3sl.ufpr.br/}}\cite{simmc} é um projeto de monitoramento de políticas públicas do MCTIC (Ministério da ciência, tecnologia, inovações e comunicações). Esse projeto realiza o monitoramento de mais de seis mil pontos de presença, coletando informações de uso de rede a cada cinco minutos e informações de inventário diariamente. Essa coleta iniciou em 2014 e todos os dados coletados desde então estão disponíveis publicamente.

O LDE (Laboratório de Dados Educacionais) \footnote{LDE: \url{http://dadoseducacionais.c3sl.ufpr.br/}} \cite{lde} é um projeto financiado pelo MEC (Ministério da Educação), que produz indicadores educacionais com séries históricas, desde o ensino Fundamental até o Ensino Superior, a partir de microdados abertos disponíveis no portal do INEP.

Ambos projetos possuem portais WEB que permitem a visualização de indicadores e realização de consultas simples, porém, o conjunto de indicadores é específico, e é difícil de realizar consultas \emph{ad-hoc} e com cruzamento de informações, que esteja fora do escopo dos indicadores disponíveis.

Diferente do \simmc{}, o LDE não é uma fonte geradora de dados. Embora os dados do LDE venham de outras fontes, o projeto realiza um importante trabalho de tratamento e organização dos dados educacionais. Por essa razão foi escolhido adicionar os dados previamente tratados pelo LDE, ao invés de utilizar os dados originais, evitando retrabalho.

Vale ressaltar que mesmo que os projeto \simmc{} e LDE tenham disponibilizado dados já tratados, para inserção desses dados no repositório foram necessários tratamentos adicionais e a realização da integração dos dados como será apresentado nas seções a seguir.

A Tabela \ref{Tab:relations} apresenta informação sumarizada sobre os dados do repositório. São apresentadas: a quantidade de tabelas existente no repositório, a quantidade de colunas de cada tabela e a quantidade de registros em cada tabela.


\begin{table}[H]\scriptsize
    \centering
    \caption{Bases abertas integradas}
    \begin{tabular}[H]{clcc|clcc}
    
        \hline
        \multicolumn{2}{c}{Tabelas}   & Colunas     & Registros    & \multicolumn{2}{c}{Tabelas}     & Colunas     & Registros \\
        \hline

        \multirow{8}{*}{\parbox{1cm}{LDE Ensino Superior}}    & aluno\_ens\_superior & 128    & 81.813.362      & \multirow{7}{*}{\parbox{1cm}{LDE Geral }}       & cidade                & 6     & 5570         \\
                                & curso\_ens\_superior & 132    & 262.691        &                        &  estado                & 4     & 27       \\
                                & docente\_ens\_superior & 52     & 3.095.302   &                           &  familias\_cadunico & 32 & 30.009.293   \\
                                &fies & 55 & 85.268.278    &                           & ibge\_pib & 8 & 27.835  \\
                                & ies\_ens\_superior      & 51 & 19.137   &                           &  pessoas\_cadunico & 36 & 90.419.338    \\
                                & localoferta\_ens\_superior & 17 & 1.396.990 &                           & pnad & 56 & 6.015.874  \\
                                & ocde\_ens\_superior & 9 & 1.162    &                           &  regiao                & 2     & 5   \\
                                & prouni                & 15    &1.069.600      &                       &  \\ \hline

        \multirow{6}{*}{\parbox{1cm}{LDE Ensino Básico}}  & aluno                 & 106   & 272.597.280    & \multirow{3}{*}{SIMMC}    & fnu                   & 9     & 2.046.110.236\\ 
                                & formação\_superior       & 6     & 255             &                           & localizacao\_ponto    & 3     & 6732  \\
                                & escola                & 185   & 1.375.322     &                               & ponto                 & 8     & 18424  \\
                                & professor & 146   & 58.190.497         &                                 \\
                                &turma                 & 95    & 12.494.116        &                                \\
                                & instituição\_superior& 8     & 6002         &                                \\
        \hline
    \end{tabular}
    \label{Tab:relations}
\end{table}

O conjunto de dados armazenados foi extraído das seguintes fontes abertas e não integradas:
\begin{itemize}
    \item SIMMCTIC \footnote{Fonte original: SIMMCTIC: \url{http://dadosabertos.c3sl.ufpr.br/simmc/}}
    \item Microdados do Censo da Educação Superior \footnote{Fonte original: Microdados INEP: \url{http://portal.inep.gov.br/microdados}}
    \item Microdados do Censo Escolar \footnote{Fonte original: Microdados INEP: \url{http://portal.inep.gov.br/microdados}}
    \item FIES \footnote{Fonte original: Indicadores sobre o FIES: \url{http://dados.gov.br/dataset/fies-fundo-de-financiamento-estudantil}}
    \item PROUNI \footnote{Fonte original : Dados Abertos do MEC: \url{http://dadosabertos.mec.gov.br/prouni}}
    \item Subconjunto da PNAD Continua \footnote{Fonte original: IBGE - PNAD continua: \url{https://www.ibge.gov.br/estatisticas/sociais/trabalho/9173-pesquisa-nacional-por-amostra-de-domicilios-}\\\url{continua-trimestral.html?=&t=downloads}}
    \item Base Desidentificada do CadUnico \footnote{Fonte original: Ministério da Cidadania: \url{https://aplicacoes.mds.gov.br/sagi/portal/index.php?grupo=212}}
\end{itemize}



\section{Armazenamento \& Disponibilização}\label{sec:armazenamento}

Para que os dados pudessem ser acessados de forma eficiente foi utilizado o SGBD MonetDB\footnote{\url{http://monetdb.org/}} como ferramenta de armazenamento dos dados. O MonetDB é um banco de dados relacional com armazenamento orientado a colunas. A escolha desse SGBD deve-se ao fato de dos dados inseridos serem estruturados e devido ao enfoque de consultas do tipo OLAP na utilização do repositório. 

Para disponibilizar os dados foi utilizada uma ferramenta chamada \blendb{}\footnote{Código fonte: \url{https://gitlab.c3sl.ufpr.br/c3sl/blendb}}. Essa ferramenta permite que o banco de dados seja acessado através de uma API RESTful, utilizando uma linguagem de consulta simplificada. A principal motivação para utilizar essa ferramenta é a possibilidade de permitir a realização de consultas diretamente no repositório, sem a necessidade de realizar uma cópia local do repositório.

A capacidade de realizar consultas no repositório permite a disponibilização de dados pré-agregados, isto é, redução da granularidade dos dados. Dados pré-agregados são menos precisos, mas possuem menor volume, o que permite que sejam processados e transferidos mais rapidamente.

Considere a base de dados de tráfego de rede do projeto \simmc{}, onde existe um registro para cada contato que cada um dos pontos já realizou. Para realização de uma análise em nível municipal e diário, uma base que contenha um registro por cidade e por dia seria igualmente útil, porém muito mais eficiente. Pensando na geração de uma cópia local da base, ao invés de transferir 2 bilhões de registros pela rede, apenas algumas milhares de registros precisariam ser transferidos, economizando armazenamento local de banda de rede.


Para obter algum dado do repositório é necessário realizar uma consulta utilizando a linguagem simplificada da ferramenta \blendb{}. Essencialmente, a consulta consiste da seleção dos atributos desejados, divididos em dois grupos, métricas e dimensões. As métricas representam os dados que serão pré-agregados e as dimensões a granularidade do dado. Se a granularidade informada na consulta for menor que a dos dados originais, o dado fornecido será pré-agregado. A base completa pode ser recuperada usando todas as dimensões em uma determinada requisição.



Um exemplo de consulta na base é mostrado a seguir: nessa consulta é retornado a quantidade de pontos de presença ativos atualmente no projeto \simmc{}, média do PIB em 2014, população em 2014, quantidade de instituições de ensino superior em 2017 e quantidade de escolas em 2017 por região do Brasil. Como a base é aberta, a utilização de todas as consultas apresentadas neste artigo pode ser colocada diretamente em um navegador WEB para obtenção da resposta.

 
 \vfill

\begin{small}
\begin{verbatim}

  http://biod.c3sl.ufpr.br/api/v1/data?metrics=
  met:count:ponto:id,met:avg:ibge:pib,met:sum:ibge:populacao,
  met:count:es:instituicao:id,met:count:escola:id
  &dimensions=dim:regiao:nome&filters=dim:ponto:ativo==t;
  dim:ibge:censo:ano==2014;dim:es:instituicao:censo:ano==2017;
  dim:escola:censo:ano==2017&sort=met:avg:ibge:pib&format=csv 

\end{verbatim}
\end{small}

A listagem de métricas e dimensões disponíveis na base atua como um dicionário de dados, contendo além de uma descrição sobre o significado do dado, informações técnicas como tipo de dados e identificador de acesso (nome) no repositório de dados. Os dados contidos no repositório hoje totalizam 685 métricas e 903 dimensões. As listas de métricas\footnote{Lista de métricas: \url{\api/metrics?format=csv}} e dimensões\footnote{Lista de dimensões: \url{\api/dimensions?format=csv}} também podem ser obtidas pela API da ferramenta \blendb{}.

Além da linguagem de consulta, o \blendb{} também possui opções que permitem definir como o resultado da consulta será exportado. Atualmente os resultados podem ser exportados como um documento JSON, um formato adequado para que páginas WEB consumam os dados, e como um arquivo CSV (parâmetro \textit{format}), ideal para a geração de uma cópia local da base, utilização por linguagens de programação como R e Python e utilização de ferramentas como Pandas e LibreOffice. Quando a resposta é exportada em CSV, ainda pode-se escolher os seguintes caracteres como separadores: vírgula, ponto-e-vírgula e tabulação.

Como o enfoque desse artigo é apresentar a base de dados, os dados que ela possui e as possibilidades e oportunidades de uso dessa base e não nas ferramentas utilizadas, detalhes sobre a ferramenta \blendb{} não são apresentados. Entretanto, mais informações sobre como a ferramenta e como utilizá-la para realizar consultas pode ser encontrada na documentação da ferramenta e na página WEB do repositório.
  

\section{Coleta e reconstrução}\label{sec:coleta}
Todos os dados disponibilizados no repositório de dados integrados foram obtidos de outras bases de dados abertos, conforme explicitado na seção \ref{sec:conteudo}. A principal contribuição desse repositório não são novos dados, mas sim uma referência única onde diversos dados diferentes podem ser encontrados de forma integrada.


Como cada entidade utiliza seu próprio padrão para disponibilizar os dados, foi necessário realizar um tratamento nos dados antes de adicionar os dados ao repositório.

O repositório de dados integrados pode ser construído automaticamente a partir dos dados das bases originais. Com exceção dos dados em si e da ferramenta \blendb{}, todos os componentes necessários para construir a base foram colocados em um repositório \emph{git} distribuído livremente\footnote{ \url{https://gitlab.c3sl.ufpr.br/simmctic/biod/biod-database}}. Esse repositório contém os \emph{scripts} necessários para realizar a construção da base e os arquivos de configuração da ferramenta \blendb{}.

Esse conjunto de \emph{scripts} é utilizado para realizar as seguintes tarefas: criar o esquema do banco de dados no SGBD MonetDB, inserir os dados tratados no banco de dados, copiar a configuração da ferramenta \blendb{} para local adequado. As ferramentas necessárias para a obtenção dos dados a partir das fontes originais não se encontra nesse repositório. A utilização dos \emph{scripts} desse repositório assumem que os dados já foram obtidos. O processo de extração dos dados a partir das fontes originais será descrito a seguir. Esse conjunto de \emph{scripts} também pode ser utilizado para inserir dados no SGBD PostgresSQL. 

Os dados do projeto \simmc{} são disponibilizados de forma aberta\footnote{Dados abertos SIMMCTIC: \url{http://dadosabertos.c3sl.ufpr.br/simmc/}} diariamente~\cite{simmcdsw2017}. Todo dia, um novo arquivo com os dados do monitoramento do dia anterior é disponibilizado. Existe um diretório para cada política pública monitorada, e dentro de cada diretório existem dois tipos de arquivo, \texttt{uso-de-rede} e \texttt{inventario}. Todos os arquivos do mesmo tipo foram combinados em uma única tabela, independente da política pública. A combinação dos dados de \texttt{uso-de-rede} gera a tabela \textbf{fnu} descrita na Tabela\ref{Tab:relations}. Os demais dados das tabelas extraídas do \simmc{} podem ser obtidas nos arquivos do diretório raiz.

Os demais dados foram obtidos através do Laboratório de Dados Educacionais (LDE). O LDE é um projeto que busca reunir em uma única plataforma dados educacionais de diversas fontes como IBGE e INEP. Além de coletar esses dados de suas fontes originais, o projeto também faz o tratamento dos dados e re-disponibilização dos dados já tratados. O tratamento dos dados é realizado através de uma ferramenta especializada de extração e integração de dados escrita em \textit{Python}, chamada \textit{HOTMapper} \cite{hotmapper}. A ferramento define mapeamentos de todas as tabelas e colunas das bases originais, em CSV, por ano, para uma única base integrada.

O \simmc{} publica dados novos diariamente. Já as demais fontes possuem periodicidade variada, podendo ser disponibilizadas anualmente, trimestralmente, ou seguindo periodicidade própria. Para manter o repositório de dados estável escolheu-se atualizar os dados de repositório anualmente, portanto, os dados do projeto \simmc{} são apenas os dados obtidos até o último dia de 2018, mesmo que na fonte existam dados mais recentes.

\section{Desafios, limitações e dificuldades}\label{sec:desafios}

Dentre os principais desafios na manutenção desse repositório estão a dificuldade de automatizar o processo de integração e na garantia de consistência dos dados. A ausência de um padrão comum nas bases de dados adicionadas, faz com que para cada nova base seja necessário um tratamento especializado, o que dificulta a utilização de um único conjunto de técnicas e ferramentas para a realização da integração. 

Outra dificuldade está na garantia de consistência entre as bases. Existem casos onde diferentes fontes divergem em relação ao mesmo dado. O problema se agrava quando cada uma das fontes contém parte dos dados, o que torna mais difícil a resolução desse tipo de conflito. O problema não reside apenas na resolução do conflito mas também em sua detecção. A existência desse tipo de conflito pode acabar por gerar análises errôneas que divergem da realidade. 

Além das dificuldades causadas pelos dados em si, ainda existem alguma limitações geradas pelas tecnologias utilizadas para manter e disponibilizar a base. A ferramenta \blendb{} permite a execução de consultas no repositório sem a necessidade de gerar uma cópia local da base, entretanto, a linguagem de consulta da ferramenta é mais limitada que a tradicional linguagem SQL, isto é, nem todas as consultas possíveis em SQL podem ser realizadas pela ferramenta. Mesmo com essa limitação acredita-se que o conjunto de operações possíveis é expressivo o suficiente para a realização de análises simples e para reduzir a granularidade dos dados e posteriormente realizar uma análise mais elaborada localmente. Além disso, essa limitação na expressividade da linguagem restringe a quantidade de junções possíveis, o que pode evitar junções impraticáveis que poderiam causar demora de resposta, ou até queda do serviço.

A base também está limitada a quantidade de informações que foi possível extrair de cada uma das fontes. A periodicidade de cada um dos dados não é a mesma, isso faz com que possam existir lacunas nos dados, dificultando análises que correlacionam dados de diferentes fontes. 


\section{Usos, Oportunidades e Contribuição}

Os dados contidos no repositório geram muitas oportunidades de análise devido ao conteúdo e tamanho da base. Em relação ao conteúdo, a base permite análise da educação e de políticas públicas em nível nacional e como as bases estão integradas, a análise pode correlacionar essas informações. Em relação ao tamanho da base, a quantidade e variedade de dados disponíveis permitem análises puramente estruturais como verificação de dados espúrios, análise de desempenho da consulta, normalização e desnormalização dos dados, entre outras.

Além dos usos dos dados em si, também existe a oportunidade de trabalhar com a integração de novos dados. Como foi previamente mencionado, existem diversas bases de dados abertos disponíveis, porém não mapeadas. Como o objetivo do projeto \emph{Blended Integrated Open Databases} é integrar a maior quantidade de bases de dados possível, a integração de outras bases com o repositório atualmente existente é uma grande oportunidade. O repositório de dados integrados pode ser um dos primeiros passos para a construção de uma base da dados integrados construída de forma colaborativa. 

Tendo o mapeamento de integração no formato da ferramenta \blendb{} e os dados tratados, as ferramentas de automatização de construção do repositório poderiam ser usadas para adicionar mais dados ao repositório. Embora a realização de tratamento e integração de dados não seja uma tarefa trivial, acredita-se que a forma como o repositório foi construído facilite a adição de novos dados. O repositório do projeto \emph{Blended Integrated Open Data} construído de forma colaborativa poderia ser utilizado como fonte central de distribuição de dados abertos, já integrados.


\section{Conclusão}\label{sec:conclusao}

O objetivo do repositório integrado de dados abertos é servir como uma fonte centralizada de dados abertos, disponibilizado de forma integrada. Esse repositório foi construído a partir de outras bases de dados abertas e a principal contribuição do repositório está na integração dessas bases e na distribuição de dados pré-agregados, e não no tratamento desses dados.

Atualmente o repositório é formado pela combinação dos dados obtidos de diversas fontes. Embora o repositório já possua uma grande variedade de dados, espera-se que cada vez mais bases de dados sejam integradas ao repositório. Planeja-se adicionar base de gastos da união disponibilizadas pelo portal da transparência da união \footnote{\url{http://www.portaltransparencia.gov.br}}, integrando-a com os dados já disponíveis no repositório.

Embora existam algumas limitações impostas pelo volume dos dados e pela ferramenta de acesso aos dados espera-se que essa base seja utilizada amplamente. Espera-se que com utilização mais intensa desses dados não apenas gere resultados interessantes mas ajude na superação das dificuldades previamente apresentadas, seja na detecção de inconsistências na base, no enriquecimento do repositório através de mais dados ou na proposta de melhorias nas tecnologias utilizadas para manter o repositório de dados.


\bibliographystyle{sbc}
\bibliography{sbc-template}

\end{document}